\documentclass[twoside,fleqn]{article}
\usepackage{espcrc2}
\usepackage{amssymb,amsmath}

\usepackage{graphicx}

\newcommand{\bra}{\langle}
\newcommand{\ket}{\rangle}

\newcommand{\om}{\omega}
\newcommand{\dl}{\delta}
\newcommand{\mhinv}{\,m_H^{-1}}
\newcommand{\meff}{m_{\text{eff}}}
\newcommand{\mueff}{\mu_{\text{eff}}}

\newlength{\colw}
\newcommand{\cpaper}{\cite{Skullerud:2003ki}}
\newcommand{\ccsId}{\cite{Smit:2002yg}}

\title{Particle distributions in electroweak tachyonic 
preheating\thanks{Talk presented by J.-I.~Skullerud}}

\author{Jon-Ivar Skullerud, Jan Smit and Anders Tranberg\address{ITF,
        University of Amsterdam, Valckenierstraat 65, 1018 XE
        Amsterdam, The Netherlands}}
\begin{document}
%%%% hack to get flush-left equations with zero indent in amsmath
\makeatletter \@mathmargin = 0pt \makeatother
%%%%
%\bibliographystyle{h-physrev4}

\begin{abstract}
We consider the out-of-equilibrium (quasi-) particle number
distributions of the Higgs and W-fields during electroweak tachyonic
preheating. We model this process by a fast quench, and perform
classical real-time lattice simulations in the SU(2)--Higgs model in
three dimensions.  We discuss how to define particle numbers and
effective energies using two-point functions in Coulomb and unitary
gauge, and consider some of the associated problems.  After an initial
exponential growth in effective particle numbers, the system
stabilises, allowing us to extract effective masses, temperatures and
chemical potentials for the particles.
\end{abstract}

\maketitle

\section{INTRODUCTION}

In recent scenarios of electroweak baryogenesis
\cite{Garcia-Bellido:1999sv,Krauss:1999ng} the electroweak transition
is assumed to have taken place at low (zero) temperature shortly after
inflation, by the effective mass-squared parameter of the Higgs field
changing sign from positive to negative (`tachyonic').  An important
issue in this scenario is the time needed for the system to reach
approximate thermalisation, and the resulting effective temperature.
The temperature must be low enough and the thermalisation sufficiently
rapid that sphaleron transitions that can wash out the generated
baryon asymmetry are prevented.  One of the main aims of this study is
to determine the effective temperature.

We employ the classical approximation, which allows us to study fields
far from equilibrium nonperturbatively by numerical simulation.  In
our case, this approximation should be justified \ccsId\ since the
instability resulting from the change of sign in the mass-squared
parameter leads to exponentially growing, and hence large, occupation
numbers.  %Confirming this will be an important test of the consistency
%of our approach.  
The full results are presented in \cpaper.

\section{PARTICLE DISTRIBUTIONS}

The definition of particle numbers and energies in an interacting
field theory out of equilibrium is not unique.  Furthermore, in a
non-abelian gauge theory the two-point correlators used to define them
may be gauge dependent.  Here, the natural choice of gauge for
the Higgs doublet $\phi$ is the unitary gauge, where it only has one
non-zero real component $\phi=(0,h/\sqrt{2})^T$.  For the gauge
fields, we will study the particle distribution in both the unitary
gauge and the Coulomb gauge $\partial_i A_i=0$.

%A natural way of defining a position-dependent distribution function
%$n(\vec{x},\vec{k},t)$ is to consider a region $R(\vec{x})$ of size
%$B$ around the position $\vec{x}$, and restrict fourier integrals to
%this region.  This leads to an intrinsic uncertainty in position and
%momenta, given by the size of the region.  For a homogeneous system,
%we may then average over all space.  This can be shown \cpaper\ to be
%equivalent to averaging the two-point correlation functions obtained
%by fourier transforms on the entire volume, over nearby points in
%momentum space, with a weight function given by the size and shape of
%the region $R$.  In practice, we implement this by `binning' all
%momenta with nearby absolute values, corresponding to spherical shells
%in position space.

We use a method \cite{Aarts:1999zn,Salle:2000hd} where effective
particle numbers and energies are determined selfconsistently, in
analogy with the free-field quantum correlators.  For the Higgs field
(with $\pi_h=\dot{h}$),
they are defined as 
\begin{align}
n^H_k &\equiv
 \sqrt{\bra h(\vec{k})h(-\vec{k})
  \ket_C\bra\pi_h(\vec{k})\pi_h(-\vec{k})\ket_C} \, ,\\
\om^H_k &\equiv
 \sqrt{\frac{\bra\pi_h(\vec{k})\pi_h(-\vec{k})\ket_C}
      {\bra h(\vec{k})h(-\vec{k})\ket_C}} \, ,
\end{align}
where $\bra\cdots\ket_C$ is the connected two-point function given by
$\bra A B\ket_C \equiv \bra AB\ket - \bra A\ket\bra B\ket$, and we
have suppressed the common time coordinate $t$.

The gauge field correlators $\bra A^a_i(k)A^b_j(-k)\ket =
\dl^{ab}C^{AA}_{ij}(\vec{k})$ can be decomposed in a transverse and a
longitudinal part, as
\begin{equation}
C^{AA}_{ij}(\vec{k}) =
\Bigl(\dl_{ij}-\frac{k_ik_j}{k^2}\Bigr)D_T^A(k)+\frac{k_ik_j}{k^2}D_L^A(k)\,,
\end{equation}
and analogously for the canonical momentum field $E^a_i=F^a_{i0}$.
In the Coulomb gauge the gauge potential is purely transverse, and
$n^A_k$ and $\om^A_k$ can be defined as
\begin{equation}
n^A_k \equiv \sqrt{D^A_T(k)D^E_T(k)}\,, \qquad
\om^A_k \equiv \sqrt{\frac{D^E_T(k)}{D^A_T(k)}} \, .
\end{equation}
In the unitary gauge, the transverse $n^T_k$ and $\om^T_k$ are found
to be the same as for the Coulomb gauge, while the longitudinal
occupation numbers and mode energies, assuming the form
$\om^{L2}_k=\meff^2+k^2$, are given by \cpaper
\begin{equation}
n^L_k = \sqrt{\!D^A_L(k)D^E_L(k)}\,, \quad
\om^L_k = \meff^2\sqrt{\frac{D^A_L(k)}{D^E_L(k)}} .
\label{eq:omL}
\end{equation}

\section{RESULTS}

\begin{figure}
\includegraphics*[width=\colw]{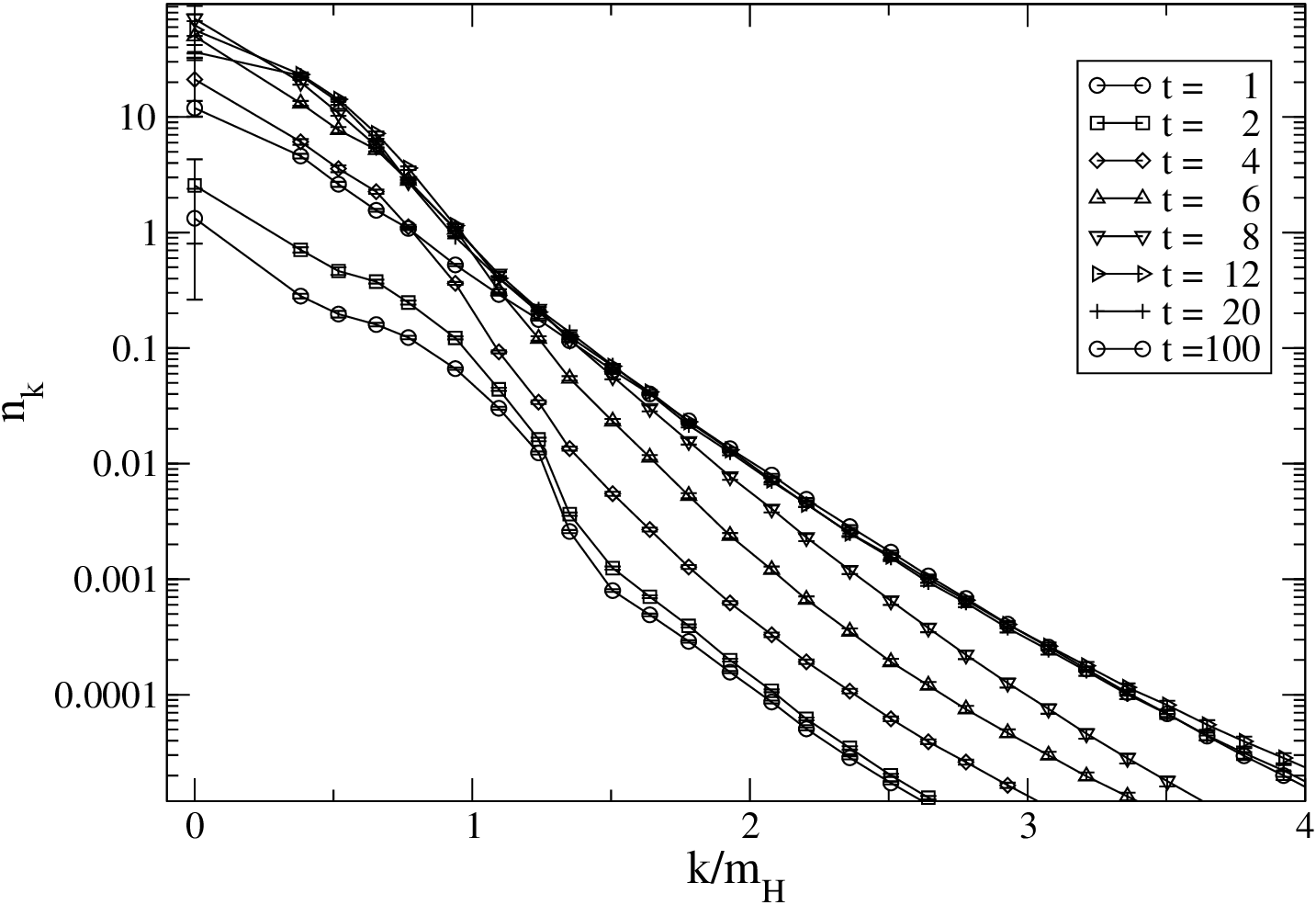}
\includegraphics*[width=\colw]{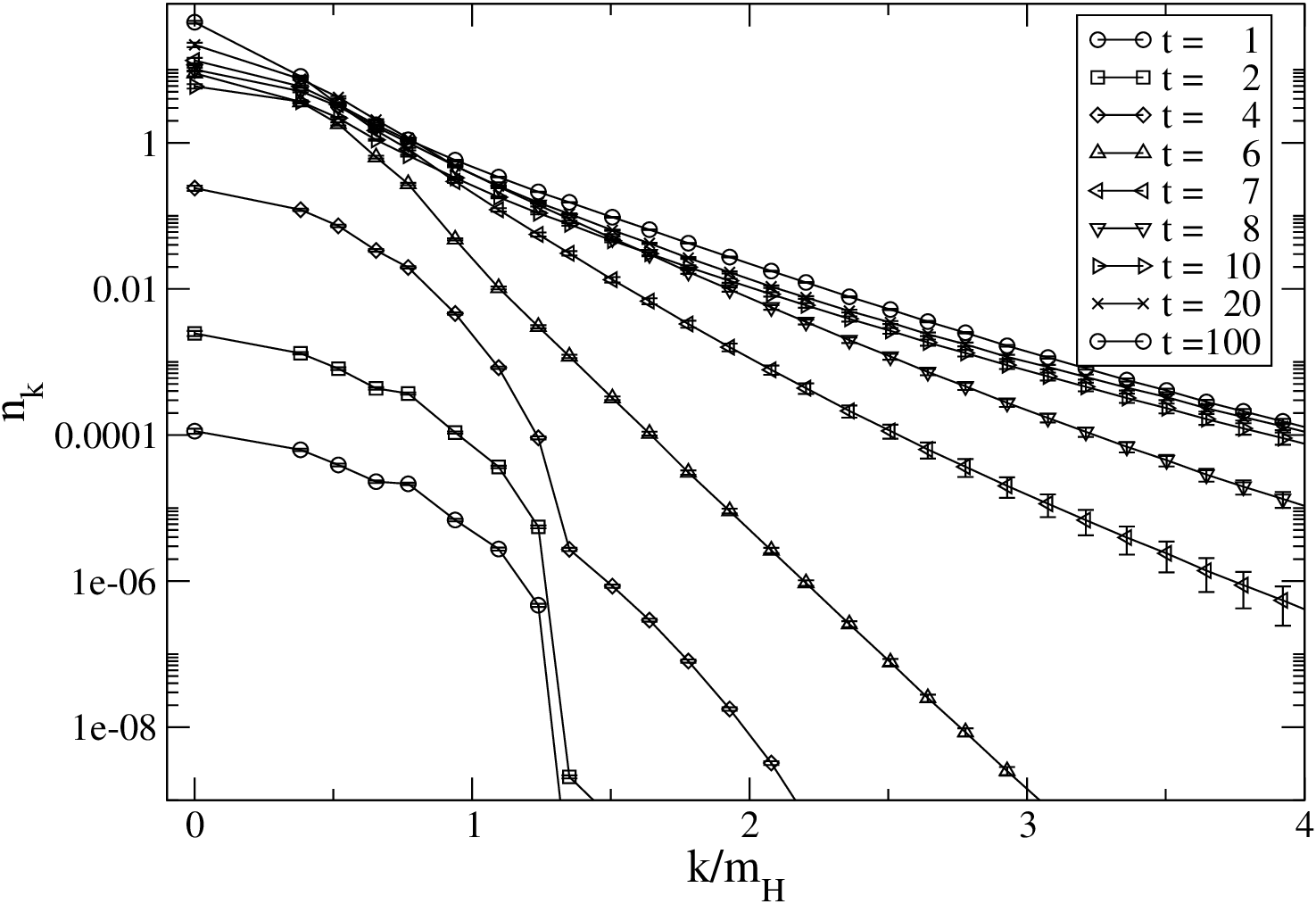}
\vspace{-1.0cm}
\caption{Higgs (top) and Coulomb-gauge W (bottom) particle
  distributions for different times.}
\label{fig:nk}
\end{figure}
We have performed simulations on a volume $L^3=21^3m_H^{-3}$ with
$g=2/3, \lambda=1/9$, giving $m_H^2/m_W^2=2$, with lattice spacing
$a=0.35\mhinv$.  We initialise the system according to the ``Just the
half'' scheme introduced and explained in \cite{Smit:2002yg}.  The
Higgs field is initialised by generating classical realisations of an
ensemble reproducing the quantum vacuum correlators. We assume that
the Higgs field is in the symmetric phase ($\bra\phi\ket=0$) with an
effective mass parameter $\mueff^2=\mu^2>0$, and approximately free
field fluctuations
\begin{equation} 
\langle\phi(k)\phi(k)^{\dagger}\rangle=\frac{1}{2\omega(k)}, \quad
\langle\pi(k)\pi(k)^{\dagger}\rangle=\frac{\omega(k)}{2},
\label{eq:higgscorr}
\end{equation}
with $\omega(k)=\sqrt{\mu^2+k^2}$.  However, we only initialise the
unstable ($|k|<|\mu|$) modes.  The gauge potential $A$
is initialised to zero, while the $E$-field is constructed to satisfy
the Gauss constraint.

We model the transition where $\mueff^2$ goes through zero as a
quench, in which $\mueff^2$ flips its sign from $\mu^2$ to $-\mu^2$
instantaneously.  42 independent realisations of the initial
conditions (\ref{eq:higgscorr}) have been generated, and the
subsequent time evolution sampled for $tm_H=1,$ 2, \ldots, 12, 20, 30,
40, 50, 100.  Nearby momenta have been averaged within `bins' of size
$\Delta=0.05a^{-1}=0.0175m_H$

Figure \ref{fig:nk} shows the particle distributions for the Higgs and
Coulomb-gauge W fields.
\begin{figure}
%\vspace{-1.0cm}
\includegraphics*[width=\colw]{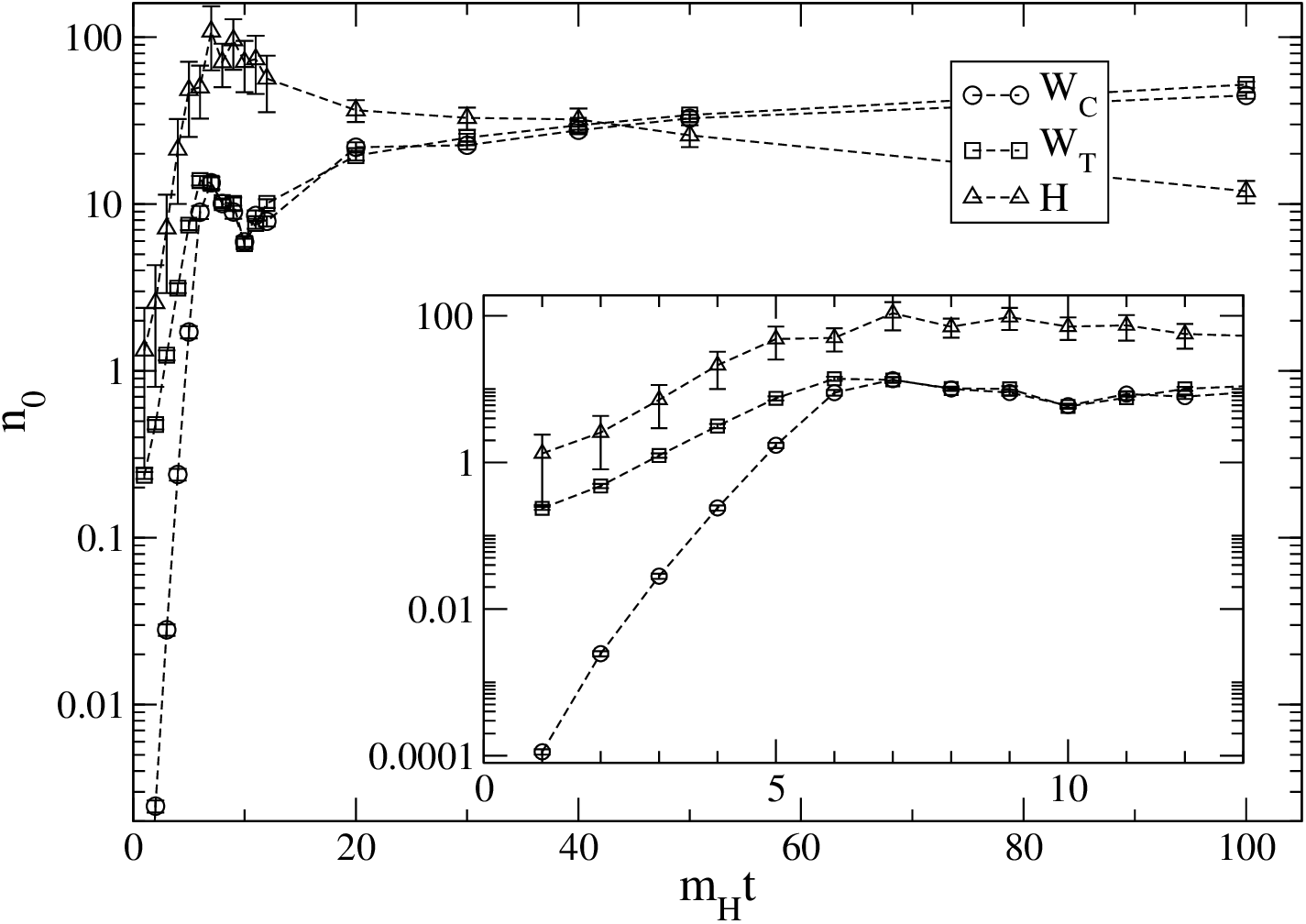}
\vspace{-1.0cm}
\caption{Particle numbers of the zero-modes as a function of time.
  $H$: Higgs; $W_T$: transverse W in unitary gauge; $W_C$: transverse
  W in Coulomb gauge.}
\label{fig:zeromod}
\end{figure}
In both cases we see the occupation numbers
of the low-momentum modes increasing exponentially up to
$t\approx6-8\,\mhinv$.  At this point the low-momentum modes start
saturating, while the high-momentum modes rapidly become populated.
After $t\approx20\,\mhinv$ the system evolves only very slowly, with
an approximately exponential distribution at late times.  A similar
behaviour is seen for the W fields in the unitary gauge.  The W
distributions in unitary and Coulomb gauge start out very different
(the unitary-gauge distribution initially being close to the Higgs
distribution) but become indistinguishable after
$t\approx40\mhinv$.  The early exponential growth and subsequent
slow evolution is illustrated in fig.~\ref{fig:zeromod}, which shows
the time evolution of the zero-mode occupation numbers.  The
exponential fall-off with momentum means that we do not expect
problems with lattice artefacts due to cutoff effects.

%\begin{figure}
%\includegraphics*[width=\colw]{Phi_disp_bw.eps}
%\vspace{-1.0cm}
%\caption{Dispersion relation for Higgs particles.}
%\label{fig:disp-phi}
%\end{figure}

%Figure \ref{fig:disp-phi} shows the dispersion relation, $\om_k^2$ as
%a function of $k^2$, for the Higgs particles.  For
%$t\lesssim20\mhinv$, exemplified here by $t=8$, there is no sensible
%dispersion relation, while for $t\gtrsim30\mhinv$ it approaches the
%form $\om^2=m_{\text{eff}}^2+ck^2$, with $c\approx1$.  The same
%behaviour is seen for the W-particles in Coulomb gauge, and
%qualitatively also for the transverse modes in the unitary gauge.  In
%the latter case, however, the particle-like behaviour is slower to
Turning to the dispersion relation, %$\om_k^2$ as a function of $k^2$,
we find qualitatively the same behaviour for the Higgs particles and
the transverse W-modes in the Coulomb and unitary gauges.  For
$t\lesssim20\mhinv$ there is no sensible dispersion relation, while
for $t\gtrsim30\mhinv$ it approaches the form $\om^2=\meff^2+ck^2$,
with $c\approx1$ for the Higgs and Coulomb-gauge W particles.  In the
unitary gauge, the particle-like behaviour is slower to emerge, and
the slope remains smaller than 1 (and increasing) at the latest times.
For the longitudinal modes, with the `inverse' definition
(\ref{eq:omL}) of the mode energy, the dispersion relation approaches
a straight line only very slowly, and we do not attempt to perform a
fit to the data.  We fit the effective energies to the form
$\om_k^2=ck^2+\meff^2$.  For the W fields, we find that the intercepts
in the two gauges are compatible and change very little from
$t=40\,\mhinv$ onwards, yielding a value $\meff\approx0.68m_H$, close
to the zero-temperature value $0.71\,m_H$.  For the Higgs fields, we
find an effective mass which is somewhat smaller than the
zero-temperature mass, although it appears to be increasing with
time.

\begin{figure}
\includegraphics*[width=\colw]{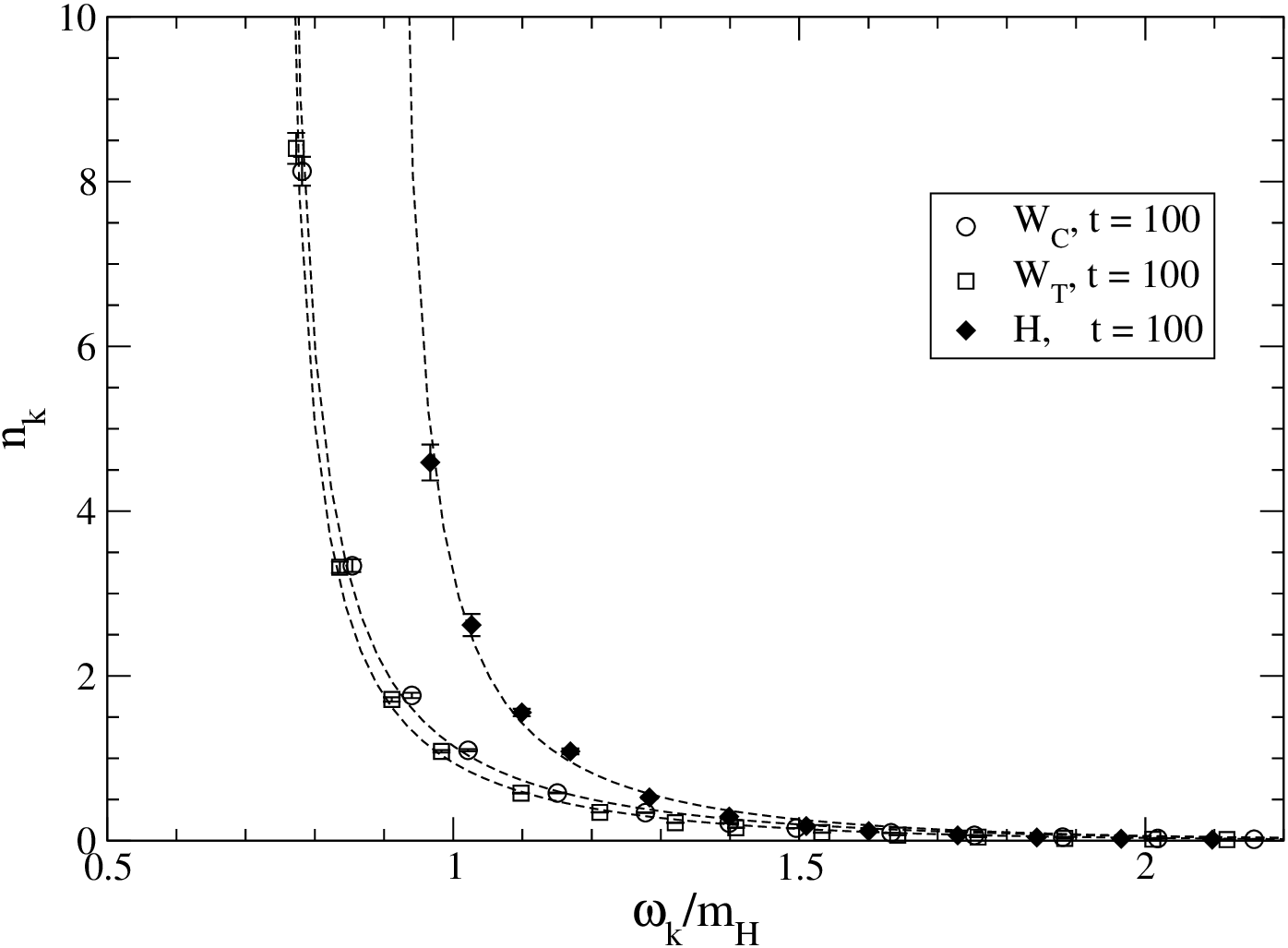}
\vspace{-1.0cm}
\caption{Higgs and W-particle numbers at $t=100\mhinv$ as a function
  of the effective energy $\om$ taken from the fitted dispersion
  relation. Also shown are the best fits to the Bose--Einstein
  distribution.  The zero-modes are outside the boundaries of the
  plot.}
\label{fig:nomk}
\end{figure}

Taking the effective energy from the fitted dispersion relation, we
then fit the particle numbers to a Bose-Einstein distribution,
$n_k = [\exp((\om_k-\mu)/T)-1]^{-1}$.
%\label{eq:be}
%\end{equation}
%
We only fit to the lowest 5 non-zero modes, for which $n_k>0.5$ and
thus the classical approximation may be expected to be valid.  The
best fits at the latest time ($t=100\mhinv$) are shown in
fig.~\ref{fig:nomk}.  We find that the effective temperature $T$ of
the Coulomb-gauge W fields stabilises from $t\approx40\mhinv$ onwards
at $T_W\approx0.42\,m_H$.  In the unitary gauge, the temperature is
lower (but increasing) due to the smaller slope in the dispersion
relation.  The effective Higgs temperature starts out higher but drops
to $T_H\approx0.4\,m_H$ at $t=100\mhinv$.  In all cases, however, a
large chemical potential --- slightly larger even than the effective
mass --- is required to describe the distribution.  This would lead
to an unphysical pole in the distribution; however, the occupation
numbers of the zero-modes (which are not included in the fits) lie
considerably below the fits, indicating a flattening of the
distribution relative to a BE form at the lowest available energies.

%\section{SUMMARY}

\section*{Acknowledgments}

This work was supported by FOM/NWO.  AT enjoyed support from
the ESF network COSLAB.

%\bibliography{nonequil}

\end{document}